\input harvmac.tex
\hfuzz 15pt
\input amssym.def
\input amssym.tex
\input epsf\def\tfig#1{{
\xdef#1{Fig.\thinspace\the\figno}}Fig.\thinspace\the\figno
\global\advance\figno by1}


\input epsf

%



\def\p{\partial}

\def\a{\alpha}
\def\b{\beta}
\def\g{\gamma}

\def\e{\epsilon}
\def\ve{\varepsilon}
\def\th{\theta}

\def\l{\lambda}

\def\th{\theta}

\def\G{\Gamma}
\def\D{\Delta}

\def\O{\Omega}

\def\ov{\over}



 %

\def\[{\left[}
\def\]{\right]}
\def\({\left(}
\def\){\right)}
\def\<{\left\langle\,}
\def\>{\,\right\rangle}


\def\inv{^{-1}}

\def\hf{{\textstyle{1\over 2}}}

 \def\frac#1#2{ {{\textstyle{#1\over#2}}}}
\def\inv{^{\raise.15ex\hbox{${\scriptscriptstyle -}$}\kern-.05em 1}}

 \def\IP{\relax{\rm I\kern-.18em P}}


%




\def\dC{C\kern-6.5pt I}

%



\chardef\tempcat=\the\catcode`\@ \catcode`\@=11
\def\cyracc{\def\u##1{\if \i##1\accent"24 i%
    \else \accent"24 ##1\fi }}
\newfam\cyrfam



\def\np#1#2#3{{Nucl. Phys.} {\bf B#1} (#2) #3}

\def\cmp#1#2#3{{Comm. Math. Phys.} {\bf #1} (#2) #3}

\def\lmp#1#2#3{{Lett. Math. Phys.} {\bf #1} (#2) #3}
\def\tmatp#1#2#3{{Theor. Math. Phys.} {\bf #1} (#2) #3}
\def\hep#1{{hep-th/}#1}

\def\encadremath#1{\vbox{\hrule\hbox{\vrule\kern8pt\vbox{\kern8pt
 \hbox{$\displaystyle #1$}\kern8pt}
 \kern8pt\vrule}\hrule}}

\def\ee{\epsilon}
\def\tphi{\tilde\phi}


\def\hepth#1{{arXiv:hep-th/}#1}

\def\np#1#2#3{{Nucl. Phys.} {\bf B#1} (#2) #3}

\def\cmp#1#2#3{{Comm. Math. Phys.} {\bf #1} (#2) #3}

\def\lmp#1#2#3{{Lett. Math. Phys.} {\bf #1} (#2) #3}
\def\tmatp#1#2#3{{Theor. Math. Phys.} {\bf #1} (#2) #3}




\lref\pog{R. Poghossian, JHEP {\bf 1401} (2014) 167; \hepth{1303.3015}.}

\lref\ppog{A. Pghosyan, H. Poghosyan, JHEP {\bf 1310} (2013) 131; \hepth{1305.6066}.}

\lref\spog{R. Poghossian,  Sov. J. Nucl. Phys. {\bf 48} (1988) 763.}

\lref\zam{A. Zamolodchikov, Sov. J. Nucl. Phys. {\bf 46} (1987) 1090.}

\lref\agt{L. Alday, D. Gaiotto, Y. Tachikawa,\lmp{91}{2010}167; \hepth{0906.3219}.}

\lref\nek{N. Nekrasov, Adv. Theor. Math. Phys. {\bf 7} (2004) 831; \hep{0206161}.}

\lref\agtr{R. Poghossian, JHEP {\bf 0912} (2009) 038; \hepth{0909.3412}.}

\lref\agtm{A. Marshakov, A. Mironov, A. Morozov, \tmatp{164}{2010}831; \hepth{09073946}.}

\lref\bbnz{A. Belavin, V. Belavin, A. Neveu, Al. Zamolodchikov, \np{784}{2007}202; \hep{0703084}.}

\lref\vb{V. Belavin, \tmatp{152}{2007}1275.}

\lref\hjs{L. Hadasz, Z. Jaskolski, P. Suchanek, JHEP {\bf 0703} (2007) 032; \hep{0611266}.}

\lref\bm{A. Belavin, B. Mukhametzhanov, JHEP {\bf 1301} (2013) 178; \hepth{1210.7454}.}

\lref\hasu{L. Hadasz, Z. Jaskolski, P. Suchanek, JHEP {\bf 0811} (2008) 060; \hepth{0810.1203}.}

\lref\bfe{A. Belavin, B. Feigin, JHEP {\bf 1107} (2011) 079; \hepth{1105.5800}.}

\lref\bbb{A. Belavin, V. Belavin, M. Bershtein, JHEP {\bf 1109} (2011) 117; \hepth{1106.4001}.}

\lref\bmt{G. Bonelli, K. Maruyoshi, A. Tanzini, JHEP {\bf 1108} (2011) 056; \hepth{1106.2505}.}

\lref\ssp{C. Crnkovic, R. Paunov, G. Sotkov, M. Stanishkov, \np{336}{1990}637.}

\lref\hjas{L. Hadasz, Z. Jaskolski, hepth{1312.4520}.}

\lref\abf{A. Belavin, M. Bershtein, B. Feigin, A. Litvinov, G. Tarnopolsky, \cmp{319}{2013}269; \hepth{1111.2803}.}

\lref\afl{V. Alba, V. Fateev, A. Litvinov, G. Tarnopolsky, \lmp{98}{2011}33; \hepth{1012.1312}.}

\lref\at{M. Afimov, G. Tarnopolsky, JHEP {\bf 1202} (2012) 036; \hepth{1110.5628}.}

\lref\gai{D. Gaiotto, JHEP {\bf 1212} (2012) 103; \hepth{1201.0767}.}

\lref\zpog{A. Zamolodchikov, R. Poghossian,  Sov. J. Nucl. Phys. {\bf 47} (1988) 929.}

\overfullrule=0pt
\Title{\vbox{\baselineskip12pt\hbox {}\hbox{}}} {\vbox{\centerline
 {On the Renormalization Group Flow}
\centerline{in Two Dimensional Superconformal Models }
  \vskip2pt
}} \centerline{Changrim Ahn$^{a,b,}$\foot{ahn@ewha.ac.kr} and Marian
Stanishkov$^{b,}$\foot{marian@inrne.bas.bg; {\it On leave of absence
from INRNE, BAS, Bulgaria}} }

 \vskip 1cm

 \centerline{ \vbox{\baselineskip12pt\hbox
{\it $^{a)}$Department of Physics, Ewha Womans University, Seoul
120-750, Korea}
 }}
\centerline{ \vbox{\baselineskip12pt\hbox {\it $^{b)}$Institute for
the Early Universe, Ewha Womans University, Seoul 120-750, Korea}
 }}


\vskip 1.5cm

\centerline{ Abstract} \vskip.5cm \noindent \vbox{\baselineskip=11pt
We extend the results on the RG flow in the next to leading order to the case of the supersymmetric minimal models $SM_p$ for $p\gg 1$. We explain how to compute the NS and Ramond fields conformal blocks in the leading order in $1/p$ and follow the renormalization scheme propsed in \pog. As a result we obtained the anomalous dimensions of certain NS and Ramond fields. It turns out that the linear combination expressing the infrared limit of these fields in term of  the IR theory $SM_{p-2}$ is exactly the same as those of the nonsupersymmetric minimal theory.  }

\Date{}
\vfill \eject

%

\newsec {Introduction}

In this paper we extend the results of the paper \pog\ to the
case of supersymmetric minimal models $SM_p$, $p\rightarrow\infty$, perturbed by the least
relevant fields. The first order corrections were already obtained a long
time ago in \spog\ . It was argued that there exists an infrared (IR)
fixed point of the renormalization group (RG) flow which coincides with the minimal
superconformal model $SM_{p-2}$. In the paper \pog\ (see also \ppog\
) the $\beta$ function, the fixed point and the matrix of anomalous
dimensions of certain fields were obtained up to the second order in
perturbation theory. That extends the famous results of
A.Zamolodchikov \zam\ . Calculation up to the second order is always a
challenge even in two dimensions. The problem is that one needs the
corresponding four-point function which is not known exactly even in
two dimensions. Fortunately, in the scheme proposed in \pog\  (which
is an extension of that proposed by Zamolodchikov in \zam\ ) one
needs the value of this function up to the zeroth order in the small
parameter $\e={2\ov p+2}$.

Basic ingredients for the computation of the
correlation functions in two dimensions are the conformal blocks.
In the last years an exact relation between the latter and the
instanton partition functions of certain $N=2$ super YM theories in
four dimensions was established by the so-called AGT correspondence
\refs{\agt\nek\agtr-\agtm}. For the $N=1$ superconformal
theories that motivated the computation of the recurrence relation
for the conformal blocks of the NS \refs{\bbnz\vb-\hjs} and Ramond
\refs{\bm-\hasu} fields of the theory. Indeed it was shown in
\refs{\bfe\bbb-\bmt} that these conformal blocks coincide with the
instanton partition functions of super YM theories in certain
spaces. With these basic ingredients in hand we computed here the
four-point functions up to the desired order.

The other difficulty arises in the regularization of the integrals. We follow here the regularization proposed in \pog\ and show that it works perfectly in our case.

One can possibly further consider the more general $SU(2)$ coset models. It was shown time ago \ssp\ that the structure constants and conformal blocks (basic ingredients for the calculation) for these theories can be obtained from just the usual minimal models by certain projected tensor product (this was recently generalized for the super-Liouville theory \hjas\ ). On that basis also a generalized AGT relation was proposed \refs{\abf\afl-\at}.

The paper \pog\ was also motivated by an alternative approach to the perturbed minimal models, the so-called RG domain wall \gai\ .
The comparison gives a perfect agreement with the perturbative calculations to the second order. Moreover it was found there that the eigenvectors corresponding to the fields of the IR CFT do not receive any $\e$ corrections and speculated to be exact.
We obtained the same result in the supersymmetric case. Moreover, the aforementioned eigenvectors are exacly the same as in the $N=0$ minimal models. One can speculate that probably this result is universal for all the coset models perturbed by the least relevant field.

This paper is organized as follows.

In Section 2 we present the $N=1$ $SM_p$ theory perturbed by the
last component of the superfield $\Phi_{1,3}$. The basic ingredients
necessary for the calculations in the second order of the perturbation
theory are presented.

In Section 3 we give some details needed for the computation of
the conformal blocks in the NS sector. We mention also the important
issue of the normalization of the fields.

Section 4 is devoted to the computation of the beta function and
the IR fixed point. It is confirmed that it coincides up to the second
order with the model $SM_{p-2}$.

The matrix of anomalous dimensions for some components of the
superfields $\Phi_{n,n\pm 2}$ and $D\bar D\Phi_{n,n}$ was computed
in Section 5. It is in perfect agreement with the first order result
in \spog\ . The same is proved also for the first component
$\phi_{n,n}$.

In Section 6 we explain how to compute the mixed conformal
blocks of the (last components of) NS and Ramond fields. They are
necessary for the calculations of the anomalous dimensions for the Ramond
fields which are also presented there. The results are again in
agreement with the conjectured RG flow to $SM_{p-2}$.

\newsec{The theory}

In this paper we consider a minimal superconformal theory $SM_p$ perturbed by the least relevant field. This theory is invariant under $N=1$ superconformal algebra with central charge $\hat c=1-{8\ov p(p+2)}$ with integer $p\ge 3$. It contains primary fields
in both NS and Ramond sectors labeled by two integers with conformal dimensions
$$
\D_{n,m}={((p+2)n-p m)^2-4\ov 8p(p+2)}(+{1\ov 16}).
$$
Here the NS (R) sector corresponds to $m-n=$even (odd) and the addition in brackets is for the Ramond fields.
The fields in the NS sector are organized in  superfields:
$$
\Phi(z,\bar z,\th,\bar\th)=\phi+\th\psi+\bar\th\bar\psi+\th\bar\th\tilde\phi.
$$
 The first (and the last) component of a spinless superfield of dimensions $\D=\bar\D$ ($\D+\hf=\bar\D+\hf$) is expressed as a product of ``chiral fields" depending on $z$ and $\bar z$, respectively. We use the same notations $\phi$ and $\tilde\phi$ below for these chiral components. If we fix the two-point function of the first component $\phi$ to one, that of the second components is $(2\D)^2$ by supersymmetry. Since it is assumed that these functions are all equal to one in the renormalization procedure, we have to normalize the second component $\tilde\phi\rightarrow {1\ov 2\D}\tilde\phi$.

We will consider the superminimal model $SM_p$ with $p\rightarrow\infty$  perturbed by the least relevant field $\tilde\phi=\tilde\phi_{1,3}$ of dimension $\D=\D_{1,3}+\hf=1-\e$, $\e={2\ov p+2}\rightarrow 0$:
$$
{\cal L}(x)={\cal L}_0(x)+\l \tilde\phi(x).
$$
It is obvious that this theory is also supersymmetric, since the perturbation can be written as a covariant super-integral over the superfield $\Phi_{1,3}$.

The two-point function of arbitrary fields up to the second order is then given by:
$$
\eqalign{
<\phi_1(x)\phi_2(0)>&=<\phi_1(x)\phi_2(0)>_0-\l\int <\phi_1(x)\phi_2(0)\tilde\phi(y)>_0 d^2y+\cr
&+{\l^2\ov 2}\int <\phi_1(x)\phi_2(0)\tilde\phi(x_1)\tilde\phi(x_2)>_0 d^2x_1 d^2x_2 +\ldots}
$$
where $\phi_1$, $\phi_2$ can be the first or the last components of a superfield or Ramond fields of dimensions $\D_1$, $\D_2$.
Since the first order corrections were considered in \spog, we will focus on the second order.

One can use the conformal transformation properties of the fields to bring the double integral to the form:
\eqn\bint{
\eqalign{
&\int <\phi_1(x)\phi_2(0)\tilde\phi(x_1)\tilde\phi(x_2)>_0 d^2x_1d^2x_2 =\cr
&=(x\bar x)^{2-\D_1-\D_2-2\D}\int I(x_1) <\tilde\phi(x_1)\phi_1(1)\phi_2(0)\tilde\phi(\infty)>_0 d^2x_1}}
where
$$
I(x)=\int |y|^{2(a-1)}|1-y|^{2(b-1)}|x-y|^{2c} d^2y
$$
and $a=2\e+\D_2-\D_1$,$b=2\e+\D_1-\D_2$, $c=-2\e$. It is well known that the integral for $I(x)$ can be expressed in terms of hypergeometric functions:
\eqn\ix{
\eqalign{
I(x)&={\pi\g(b)\g(a+c)\ov \g(a+b+c)}|F(1-a-b-c,-c,1-a-c,x)|^2+\cr
&+{\pi\g(1+c)\g(a)\ov \g(1+a+c)}|x^{a+c}F(a,1-b,1+a+c,x)|^2 }}

This form is useful for evaluating $I(x)$ near $x=0$.
Using the transformation properties of the hypergeometric functions, \ix\ can be rewritten as a function of $1-x$ and ${1\ov x}$ which is suitable for the investigation of $I(x)$ around the points $1$ and $\infty$, respectively.

It is clear that the integral \bint\ is singular. We follow the
regularization procedure proposed in \pog\ . It consists basically
in cutting discs in the two-dimensional surface of radius $l$
(${1\ov l}$) around singular points $0$, $1$ ($\infty$):
$D_{l,0}=\{x\in C,|x|<l\}$, $D_{l,1}=\{x\in C,|x-1|<l\}$,
$D_{l,\infty}=\{x\in C,|x|>1/l\}$ with $0\ll l_0\ll l<1$ where $l_0$ is the ultraviolet cut-off.
Clearly $l$ should be canceled in the calculations and should not appear in the
final result. We call the region outside these discs as $\O_{l,l_0}$ where the integration is well-defined.
 It is useful to do this integration in
radial coordinates. Since the correlation function exhibits poles
only at the points $0$ and $1$, the phase integration can be
performed by using residue theorem and the resulting rational
integral in the radial direction is straightforward. Near the
singular points one can use the OPE. In doing that it turns out that
we count twice two lens-like regions around the point $1$ so we have
to subtract those integrals. We
refer to \pog\ for the explicit formulas as well as a more
detailed explanation.

\newsec{Computation of the conformal blocks in the NS - sector}

Let us start with the correlation function that enters in the
integral \bint. The basic ingredients for the computation of the
four-point correlation functions are the conformal blocks. These are
quite complicated objects in general and closed formula were not
known. Recently it was argued that they coincide (up to factors)
with the instanton partition function of certain $N=2$ YM theories
on ALE spaces, which was proved by a recurrence
relation for  the conformal blocks \refs{\bbnz\vb-\hjs}. We need the expressions for
the first few levels conformal blocks in order to have a guess for
the limit $\e\rightarrow 0$.

The chiral components of the fields obey the OPEs: \eqn\bops{
\eqalign{ \phi_1(x)\phi_2(0)&=x^{\D-\D_1-\D_2}\sum_{N=0}^\infty x^N
C_N \phi_\D(0)\cr \tilde
\phi_1(x)\phi_2(0)&=x^{\D-\D_1-\D_2-1/2}\sum_{N=0}^\infty x^N \tilde
C_N \phi_\D(0)\cr \phi_1(x)\tilde
\phi_2(0)&=x^{\D-\D_1-\D_2-1/2}\sum_{N=0}^\infty x^N \tilde C'_N
\phi_\D(0)\cr \tilde \phi_1(x)\tilde
\phi_2(0)&=x^{\D-\D_1-\D_2-1}\sum_{N=0}^\infty x^N C'_N \phi_\D(0) }
} where $C_N$'s are polynomials of order $N$ in the generators of the
superconformal algebra $L_{-k}$ and $G_{-\a}$ ($k,\a>0$) with coefficients depending on the
dimensions $\D$, $\D_1$, $\D_2$. which we omitted)of
dimension $N$ usually called chain vectors. Here $N$ runs over all
nonnegative integers or half-integers depending on
the fusion rules of $SM_p$.

Acting by positive mode generators on the both sides of these OPEs and using the super-conformal transformation properties gives the chain equations for $L$'s:
$$
L_kC_N=(\D+k\D_1-\D_2+N-k)C_{N-k}
$$
(here $C$ is any of of the chain vectors with the corresponding
dimensions of the fields) and for $G$'s: \eqn\chr{ \eqalign{ G_k C_N
&=\tilde C_{N-k}\cr G_k\tilde C_N &=(\D+2k\D_1-\D_2+N-k)C_{N-k}\cr
G_k\tilde C'_N &=C'_{N-k}\cr G_k
C'_N&=(\D+2k\D_1-\D_2+N-k-\hf)\tilde C'_{N-k}}} for
$k>\hf$, and
$$
\eqalign{
G_\hf \tilde C'_N&=2\D_2C_{N-\hf}+C'_{N-\hf}\cr
G_\hf  C'_N&=-2\D_2\tilde C_{N-\hf}+(\D+\D_1-\D_2+N-1)\tilde C'_{N-\hf} .}
$$

There are two independent constants at the zeroth level in the OPEs \bops, the other two are expressible through them:
$$
\tilde C'_0=-\tilde C_0,\hskip1cm C'_0=(\D-\D_1-\D_2)C_0.
$$
The above chain relations could be solved order by order. As
mentioned before, in \refs{\vb,\hjs}, a recursion relation for the chain
vectors can be also found. We give here as an example and for further
use the first terms for $C'$: \eqn\cfir{ \eqalign{
C'_\hf&=-{\D+\D_1+\D_2-\hf\ov 2\D}\tilde C_0 G_{-\hf}\cr
C'_1&={\D+\D_1-\D_2\ov 2\D} C_0 L_{-1}.}}
The conformal blocks are readily obtained by the chain vectors.  Presented as
vectors in the basis of $L$'s and $G$'s, the conformal block can be
expressed as:
$$
F(\D,\D_{i})=\sum_{N=0}^\infty x^N F_N =\sum_{N=0}^\infty x^N C_N(\D,\D_3,\D_4)S_N^{-1}C_N(\D,\D_1,\D_2)
$$
where $S_N$ is the Shapovalov matrix at level $N$. What of $C_N$'s appear depends on the external fields involved.

The conformal blocks are in general quite complicated objects.
Fortunately, in view of the renormalization scheme and the
regularization of the integrals, we need to compute them here only
up to the zero-th order in $\e$. This simplifies significantly the
problem.

Once the conformal blocks are known, the correlation function of spinless fields for our $SM_p$ models is written as:
$$
\sum_n C_n|F(\D_n,\D_{i})|^2
$$
where the range of $n$ depends on the fusion rules and $C_n$ is the
corresponding structure constant. Let us stress that the various
structure constants are connected. If we call the
structure constant of three first-component operators $C$, and that of two-first
and one last-component $\tilde C$, other remaining structure constants are given by
\eqn\cnor{ \eqalign{
<\tilde\phi_1(\infty)\tilde\phi_2(1)\phi_3(0)>&=(\D_3-\D_1-\D_2)^2
C_{(1)(2)(3)}\cr
<\tilde\phi_1(\infty)\tilde\phi_2(1)\tilde\phi_3(0)>&=(\hf-\D_1-\D_2-\D_3)^2
\tilde C_{(1)(2)(3)} .} }

The structure constants $C_{(1)(2)(3)}$
and $\tilde C_{(1)(2)(3)}$ were obtained in \zpog. We have to keep
in mind also that our last components are normalized by $1/2\D$. In
what follows we compute the conformal blocks up to sufficiently high
level and then check also the crossing symmetry and the behavior
near the singular points $1$ and $\infty$.

\newsec{$\beta$-function and fixed point}

For the computation of the $\b$-function in the second order, we need
the four-point function of the perturbing field. Here we consider a
more general function
$$<\tphi(x)\tphi(0)\tphi_{n,n+2}(1)\tphi_{n,n+2}(\infty)>.
$$
There are three ``channels" (or intermediate fields) in the corresponding
conformal block: two even, corresponding to the identity and
$\phi_{1,5}$ and one odd - to $\tilde\phi$ itself. From the
procedure we explained above, we get the following expression for
this correlation function:
$$
\eqalign{
&<\tilde\phi(x)\tilde\phi(0)\tilde\phi_{n,n+2}(1)\tilde\phi_{n,n+2}(\infty)>\cr
&= \left|{(1 - 2 x + 7/3 x^2 - 4/3 x^3 + 1/3 x^4)\ov x^2 (1 - x)^2}\right|^2+{2(n + 3)\ov 3 (n + 1)}\left| {(1 - 3/2 x + 3/2 x^2 - 1/2 x^3)\ov x (1 - x)^2}\right|^2\cr &+ {(3 + n) (4 + n)\ov 18  n (1 + n)}\left| {(1 - x + x^2)\ov (1 - x)^2}\right|^2. }
$$
We checked explicitly the crossing symmetry and the $x\rightarrow 1$ limit of this function.
The function that enters the integral is obtained by the
conformal transformation $x\rightarrow 1/x$ (explicit formula is
presented below):
\eqn\npt{\eqalign{
&<\tilde\phi(x)\tilde\phi_{n,n+2}(0)\tilde\phi_{n,n+2}(1)\tilde\phi(\infty)>= \left|{(1 - 4
x + 7 x^2 - 6 x^3 + 3 x^4)\ov 3 x^2(-1 + x)^2} \right|^2\cr &
+{2(n + 3)\ov 3 (n+ 1)}\left|{ (-1 + 3 x - 3 x^2 + 2 x^3)\ov 2 x^2(-1 + x)^2 }\right|^2+ {(3
+ n) (4 + n)\ov 18  n (1 + n)}\left|{(1 - x + x^2)\ov x^2(-1 + x)^2}\right|^2.
}}
In order to compute the $\beta$-function and the fixed point to the second order,
we just have to integrate the above function with $n=1$.

The integration over the safe region far from the singularities
yields ($I(x)\sim {\pi\ov \epsilon}$):
$$
\eqalign{
&\int_{\Omega_{l,l_0}} I(x)<\tilde\phi(x)\tilde\phi(0)\tilde\phi(1)\tilde\phi(\infty)>d^2x\cr
&=-{35 \pi^2\ov 24 \ee} + {2 \pi^2\ov\ee l^2} + {\pi^2\ov 2 \ee l_0^2} -
{ 16 \pi^2 \log l\ov 3 \ee} - {8 \pi^2 \log 2 l_0\ov 3 \ee} }
$$
and we omitted the terms of order $l$ or $l_0/l$.

We have to subtract the integrals over the lens-like regions
since they  would be accounted twice. We need to expand the function
around $1$ and compute the integrals using the formulas in \pog\ .
Here is the result of that integration:
$$
{\pi^2 \ov \ee}
\(-{1\ov l^2}+{1\ov 2 l_0^2} +{61\ov 24}-{8\ov 3}\log{l\ov 2l_0}\).
$$

Next we have to compute the integrals near the singular points $0,1$ and $\infty$. For that purpose we can use the OPE of the
fields and take the appropriate limit of $I(x)$.
Near the point $0$ the relevant OPE is:
$$
\tilde\phi(x)\tilde\phi(0)=(x\bar x)^{-2(\Delta_{1,3}+\hf)}(1+\ldots)
+ \hat C_{(1,3)(1,3)}^{(1,3)}(x\bar x)^{-(\Delta_{1,3}+\hf)}(\tilde\phi(0)+\ldots).
$$
The channel $\phi_{1,5}$ gives after integration a term proportional
to $l/l_0$ which is negligible. The structure constant is the one
of the {\bf normalized second components} of a superfield. Here and below we denote the
normalized structure constant by $\hat C$.
Then, the correct value is
$$
\hat C_{(1,3)(1,3)}^{(1,3)}={(\hf -3\Delta_{1,3})^2\ov (2\Delta_{1,3})^3}\tilde C_{(1,3)(1,3)}^{(1,3)} ={2\ov \sqrt 3} - 2 \sqrt 3 \ee
$$
to the first order in $\ee$. The value of $I(x)$ near $0$ can be found
by taking the limit in \ix\ written in terms of $1/x$ (explicit form
is given in \pog).
Finally one gets:
$$
\int_{D_{l,0}\backslash D_{l_0,0}} I(x)<\tilde\phi(x)\tilde\phi(0)\tilde\phi(1)\tilde\phi(\infty)>d^2x
=-{\pi^2\ov l^2 \ee} + {8 \pi^2\ov
 3 \ee^2} -{16 \pi^2\ov\ee} + {8\ov3} {\pi^2 \log l\ov\ee}.
 $$
Since the integral near $1$ gives obviously the same result, we just need to add the above result twice.
To compute the integral near infinity, we use a relation
\eqn\infi{
<\phi_1(x)\phi_2(0)\phi_3(1)\phi_4(\infty)>=(x\bar x)^{-2\D_1}<\phi_1(1/x)\phi_4(0)\phi_3(1)\phi_2(\infty)> }
and $I(x)\sim {\pi\ov\ee}(x\bar x)^{-2\ee}$.
This gives
$$
\int_{D_{l,\infty}\backslash D_{l_0,\infty}} I(x)<\tphi(x)\tphi(0)\tphi(1)\tphi(\infty)>d^2x
=-{\pi^2\ov l^2 \ee} + {4 \pi^2\ov 3 \ee^2}  - {8 \pi^2\ov\ee} + {8 \pi^2 \log l\ov3\ee}.
 $$

Putting all together, we finally obtain the finite part of the integral:
$$
{20\pi^2\ov 3 \ee^2}-{44\pi^2\ov\ee}.
$$
Here we want to mention that we follow the renormalization scheme
proposed in \pog. Therefore we already omitted the terms
proportional to $l_0^{4\ee-2}$ which could be canceled by an
appropriate counterterm in the action.

Taking into account also the first order term (proportional to
the above structure constant and computed in\spog), we get the final
result (up to the second order) for the two-point function of the
perturbing field:
\eqn\twopt{
\eqalign{
G(x,\l)&=<\tphi (x)\tphi(0)>\cr
&=(x\bar x)^{-2+2\ee}\left[1-\l {4\pi\ov \sqrt 3}\({1\ov\ee}-3\)(x\bar x)^\ee+{\l^2\ov 2}\({20\pi^2\ov 3 \ee^2}-{44\pi^2\ov\ee}\)(x\bar x)^{2\ee}
+\ldots\right]. }}

Now we introduce a field $\tphi^g=\p_g {\cal L}$ which is normalized by $<\tphi^g (1)\tphi^g(0)>=1$.
Under the scale transformation $x^{\mu}\to t x^{\mu}$, the Lagrangian transforms to the trace of energy-momentum tensor $\Theta$,
$$
\Theta(x)=\p_t {\cal L}=\beta(g)\p_g {\cal L}=\beta(g)\tphi^g.
$$
Comparing these with the orginal bare Lagrangian where $\tphi=\p_{\l}{\cal L}$ and $\Theta=\ee\l\tphi$ lead to
the $\b$-function given by
$$
\beta(g)=\ee\l{\p g \ov\p\l}=\ee\l\sqrt{ G(1,\l)},
$$
where $G(1,\l)$ is given by \twopt\ by $x=$.
One can invert this and compute the bare coupling constant and the $\beta$-function in terms of $g$:
\eqn\bare{\eqalign{
\l&=g+g^2{\pi\ov \sqrt 3}\left({1\ov\ee}-3\right)+g^3{\pi^2\ov 3}\left({1\ov\ee^2}-{5\ov\ee}\right)+{\cal O}(g^4),\cr
\beta(g)&=\e g-g^2{\pi\ov\sqrt 3}(1-3\e)-{2\pi^2\ov 3}g^3+{\cal O}(g^4).}}
In this calculatins, we keep only the relevant terms by assuming the
coupling constant $\l$ (and $g$) to be order of ${\cal O}(\e)$.

A non-trivial IR fixed point occurs at the zero of the $\beta$-function
\eqn\fx{
g^*={\sqrt{3}\ov\pi}\e(1+\e).}
It corresponds to the IR CFT  $SM_{p-2}$ as can be seen from central charge:
$$
c^*-c=-8\pi^2\int_0^{g^*}\beta(g)d g=-4\e^3-12\e^4+{\cal O}(\e^5).
$$
The anomalous dimension of the perturbing field becomes
$$
\D^*=1-\p_g\beta(g)|_{g^*}=1+\e+2\e^2+{\cal O}(\e^3)
$$
which matches with that of the second component of the superfield
$\Phi_{3,1}^{p-2}$ of $SM_{p-2}$.

\newsec{ Mixing of the super-fields in the NS sector}

The second component of a super-field as a perturbing field guarantees the preservation of
super-symmetry along the RG flow.
The dimension which is close to ($1/2,1/2$) and the fusion rules between the super-fields $\Phi_{n,n\pm
2}$ and $D\bar D\Phi_{n,n}$ where $D$ is the covariant super-derivative
suggest that the operators mix along the RG-trajectory.
We will compute the corresponding dilatation matrix for the anomalous dimensions of the second
components while  the mixing of the first ones is a consequence of the supersymmetry.
For this purpose we compute the two-point functions and the corresponding integrals.

\subsec{ Two-point function $<\tphi_{n,n+2}(1)\tphi_{n,n+2}(0)>$ }

The corresponding function in the second order of the perturbation was
already written above \npt . The integration over the safe region
(far from the singularities) goes in the same way as before. The
result is:
$$
\eqalign{
&\int_{\Omega_{l,l_0}} I(x)<\tphi(x)\tphi_{n,n+2}(1)\tphi_{n,n+2}(0)\tphi(\infty)>d^2x={(n+2)\pi^2\ov 6 \e n l_0^2}
+{2 (1 + 2 n)\pi^2\ov 3 n\e  l^2 }\cr
&-{ 4(1 + 5 n + 2 n^2)\pi^2 \log l\ov 3 \e (n + n^2)}-{4 (1 +n)\pi^2\log{2l_0}\ov 3n\e}-{(18 + 43 n + 9 n^2)\pi^2\ov 24 \e
(n +n^2)}. }
$$
Also the integration over the lens-like regions
gives similarly:
$$
{\pi^2\ov 24 n (1 + n)\e}\left[(1 + n) (2 + n)\left({4\ov l_0^2}-{8\ov l^2}\right)+(46 + n (53 + 23 n))+32 (1 +n)^2\log {l\ov{2l_0}}\right].
$$

Taking the integral around zero should be more careful.
It turns out that one should take into account the descendents since they contribute nontrivial singular terms.
Explicit OPE is
\eqn\openp{\eqalign{
&\tphi(x)\tphi_{n,n+2}(0)=(x\bar x)^{-(\D_{1,3}+1/2)} \hat C_{(1,3)(n,n+2)}^{(n,n+2)}\tphi_{n,n+2}(0)+(x\bar x)^{\delta\D-\D_{1,3}-1} \hat C_{(1,3)(n,n+2)}^{(n,n)}\cr
&(1+{\D_{1,3}+\delta\D\ov 2\D_{n,n}}x L_{-1})(1+{\D_{1,3}+\delta\D\ov 2\D_{n,n}}\bar x \bar L_{-1})\phi_{n,n}(0),
\qquad \delta\D=\D_{n,n}-\D_{n,n+2}\ }}
(the coefficient in front of $L_{-1}$ is obtained from the chain relations \cfir). Since $L_{-1}$ acts as a derivative, we get
$$
<L_{-1}\phi_{n,n}(0)\tphi_{n,n+2}(1)\tphi(\infty)>=(\D_{n,n}+\D_{n,n+2}-\D_{1,3})<\phi_{n,n}(0)\tphi_{n,n+2}(1)\tphi(\infty)>.
$$
Let us stress again that the structure constants needed for the calculation are the ``normalized" ones:
\eqn\str{\eqalign{
\hat C_{(1,3)(n,n+2)}^{(n,n+2)}&={(\hf-\D_{1,3}-2\D_{n,n+2})^2\ov 2\D_{1,3}(2\D_{n,n+2})^2} \tilde C_{(1,3)(n,n+2)}^{(n,n+2)}
={(3 + n)^2\ov 3 (1 + n)^2}-{2 (2 + n) (3 + n)^2 \e\ov 3 (1 + n)^2}\cr
\hat C_{(1,3)(n,n+2)}^{(n,n)}&={(\D_{n,n}-\D_{1,3}-\D_{n,n+2})^2\ov 2\D_{1,3}2\D_{n,n+2}}C_{(1,3)(n,n+2)}^{(n,n)}=\sqrt{{n+2\ov 3n} }.\ } }
With the same $I(x)$, the integral corresponding to the channel $\phi_{n,n}$ becomes
$$
\eqalign{
&-{(2 + n)\pi^2\ov 3\e n l^2}+{2 (-1 + n)^2 (2 + n) (5 + n) \pi^2\ov 3 \e^2 n (1 + n)^2 (3 + n)^2}+\cr &+{(2 + n) \pi^2 (-4 (-1 + n) (-1 + 23 n + 9 n^2 + n^3) +4 (-1 + n)^2 (3 + n)^2 \log l)\ov 6 \e n (1 + n)^2 (3 + n)^2}.}
$$
The channel $\tphi_{n,n+2}$ is simpler since it is sufficient to take $I(x)$ just to order of $1$ without any descendant:
$$
{2 (3 + n)^2 \pi^2\ov 3 (1 + n)^2 \e^2}-{(3 + n)^2 \pi^2 (8 + 4 n - 2 \log l)\ov  3 (1 + n)^2 \e}.
$$
The integrals around $1$ are obviously the same, so the total contribution is twice the sum of above two terms.

Computation around infinity is almost the same as the one for the $\beta$-function if we put the correct structure constants:
$$
-{\pi^2\ov\e l^2}+ {2 (3 + n) \pi^2\ov 3 (1 + n) \e^2}-{2\pi^2((3 + n) (5 + n) - (2 n + 6)\log l)\ov 3 (1 + n) \e}.
$$
Finally, combining all the terms, we get:
$$
-{2\pi^2 (-20 - 143 n - 121 n^2 - 33 n^3 - 3 n^4)\ov
 3 n (1 + n) (3 + n)^2 \e^2}-
{2\pi^2(5 + n) (8 + 151 n + 143 n^2 + 45 n^3 + 5 n^4) \ov 3 n (1 + n) (3 + n)^2 \e}.
$$
Note that the final result is very similar with \pog\  although the various integrals differ explicitly.
This will be also the case with the next integrals.

\subsec{Function $<\tphi_{n,n+2}(1)\tphi_{n,n-2}(0)>$ }

The relevant four-point function in this case in the zeroth order of $\e$ is
$$
<\tphi(x)\tphi(0)\tphi_{n,n+2}(1)\tphi_{n,n-2}(\infty)>=
{1\ov 3} \sqrt{{(-4 + n^2)\ov n^2}}\left|{ 1\ov(1 - x)^2 }(1 - x + x^2)\right|^2.
$$
$\phi_{1,5}$ is only channel appearing here. Transforming $x\rightarrow
{1\ov x}$ and  using \infi, one obtains
$$
<\tphi(x)\tphi_{n,n+2}(1)\tphi_{n,n-2}(0)\tphi(\infty)>={1\ov 3} \sqrt{{(-4 + n^2)\ov n^2}}\left|{ 1\ov x^2(1 - x)^2 }(1 - x + x^2)\right|^2
$$
which can be inserted into \bint \ (note that this is different from \pog\ ). For the integral over the
safe region we need $I(x)$ which can be extracted from \ix:
$$
I(x)=-{4\pi\ov (n^2-4)\e}.
$$
Then, the integral becomes
$$
\eqalign{
&\int_{\Omega_{l,l_0}} I(x)<\tphi(x)\tphi_{n,n+2}(1)\tphi_{n,n-2}(0)\tphi(\infty)>d^2x=\cr
&=-{4 \pi^2\ov 3 \e n \sqrt{-4 + n^2} l^2}
-{2 \pi^2\ov 3 \e n \sqrt{-4 + n^2} l_0^2}
+{\pi^2 (9 + 16 \log{(2 l l_0)})\ov 6 \e n \sqrt{-4 + n^2}}. }
$$
Expanding around $1$ and taking the integrals in the lens-like regions gives
$$
-{\pi^2 (23 - {8\ov l^2} + {4\ov l_0^2} + 16 \log{{l\ov 2l_0}})\ov
 6 e n \sqrt{-4 + n^2}},
$$
which should be subtracted.

The integral around the point $0$ is very similar to that in the previous section. The only difference is that we take $\D_{n,n-2}$ instead of $\D_{n,n+2}$ in \openp. Also, in the computation of the appropriate approximation of $I(x)$ we have to expand the hypergeometric functions for the channel $\phi_{n,n}$ up to order $x$. The computation for the channel $\tphi_{n,n-2}$ is the same as above. Finally, we need the structure constant:
$$
\hat C_{(1,3)(n,n-2)}^{(n,n)}={(\D_{n,n}-\D_{1,3}-\D_{n,n-2})^2\ov 2\D_{1,3}2\D_{n,n-2}}C_{(1,3)(n,n-2)}^{(n,n)}=\sqrt{{n-2\ov 3n}}.
$$
At the end we get:
$$
\eqalign{
&\int_{D_{l,0}\backslash D_{l_0,0}} I(x)<\tphi(x)\tphi_{n,n+2}(1)\tphi_{n,n-2}(0)\tphi(\infty)>d^2x=\cr
&={4 \pi^2\ov 3 \e^2  n (-9 + n^2) \sqrt{-4 + n^2}} \left[10  + \e {(-9 + n^2)\ov l^2} - 2 \e (1 + n^2) -
   2 \e  (-9 + n^2) \log l\right]. }
$$
In principle one should compute also the integral around $1$. Just
as \pog, it turns out to be the same as that around
$0$. So it is enough to take the above result twice. Also, the
integral around $\infty$ is not singular here and can be neglected.
Collecting the integrals computed above gives
$$
{80 (1 - 2 \e) \pi^2\ov 3 \e^2 n (-9 + n^2) \sqrt{-4 + n^2}}.
$$
Again, the finite result is similar to that of \pog\ even though individual integrals are
different.

\subsec{Function $<\phi_{n,n}(1)\tphi_{n,n+2}(0)>$ }
The integration over the safe region is
$$
\eqalign{
&\int_{\Omega_{l,l_0}} I(x)<\tphi(x)\phi_{n,n}(1)\tphi_{n,n+2}(0)\tphi(\infty)>d^2x=\cr
&={8 \e\pi^2\ov 3 (5 + n)}\sqrt{{
 2 + n\ov n}}  \left[(-5 + 3 n) \log l + (1 + n) \log{2 l_0}\right] }
$$
where the integrand is given by
$$
<\tphi(x)\phi_{n,n}(1)\tphi_{n,n+2}(0)\tphi(\infty)>={2\ov 3}\sqrt{{n+2\ov n}}|x|^{-2}.
$$
with  $I(x)$ given above.
The integration over the lens-like region is similarly given by
$$
-{8 \e (1 + n)\pi^2\ov 3 (5 + n)}  \sqrt{{
 2 + n\ov n}} \log {{l\ov 2 l_0}}.
$$

The OPE needed for computation around $0$ was written above \openp. One has to arrange the corresponding dimensions and structure constants from \str. So the contribution from the region near $0$ is:
$$
\eqalign{
&\int_{D_{l,0}\backslash D_{l_0,0}} I(x)<\tphi(x)\phi_{n,n}(1)\tphi_{n,n+2}(0)\tphi(\infty)>d^2x=\cr
&-{4 (-1 + n) \sqrt{{2 + n\ov n}} \pi^2\ov 3 (3 + n) (5 + n)}[1+(n + (n+3) \log l)\e]
-{2(n+3)\pi^2 \sqrt{{2 + n\ov n}}\ov 3 (5 + n)}   (1 + 3 \e + 2 \e \log l). }
$$
Surprisingly the computation around the point $1$ again gives a result identical to that around $0$.
So we have to add again twice the above contribution.

To compute the contribution from the region near $\infty$, we
perform again the $x\rightarrow 1/x$ map \infi. The necessary
structure constants are already written above and we take the
appropriate (up to $\e^2$) approximation for $I(x)$. The result is:
$$
\eqalign{
&\int_{D_{l,\infty}\ D_{l_0,\infty}}I(x)<\tphi(x)\phi_{n,n}(1)\tphi_{n,n+2}(0)\tphi(\infty)>d^2=\cr
&=-{8 (-3 + n)\pi^2\ov 3(n+5)}\sqrt{{n+2\ov n}}
(1+ (2 + n + 2\log l ) \e). }
$$
Combining all the terms, we get
$$
-{4 (-1 + n)\pi^2\ov 3 (3 + n) (5 + n)}\sqrt{{n+2\ov n}}[11 + 3 n + \e (1 + n) (9 + 2 n))].
$$

\subsec{Function $<\phi_{n,n}(1)\phi_{n,n}(0)>$ }

Finally we need the function $<\tphi(x)\phi_{n,n}(1)\phi_{n,n}(0)\tphi(\infty)>$.
This function happens to coincide exactly with the one reported in \pog\ .
Therefore almost all integrals are the same. The only exception  is
the integral around $\infty$ due to a different structure constant:
$$
\hat C_{(1,3)(n,n)}^{(n,n)} \hat C_{(1,3)(1,3)}^{(1,3)}={(-1 + n^2)  \e^2\ov 6 n }(1-2\e).
$$
With this, our final result is
$$
{(-1 + n^2)\pi^2 \ov 6}(1 + \e)
$$
which is slightly different from \pog.

Since the dimension of the first component $\phi_{n,n}$ is close to zero, it doesn't mix with other fields.
Therefore, we need to compute only its anomalous dimension.
Taking into account also the first order
contribution, the final result for the two-point function is:
$$
\eqalign{ G_n(x,\l)=<\phi_{n,n}(x)\phi_{n,n}(0)>&=(x\bar
x)^{-2\D_{n,n}}\left[1 - \l \left({\sqrt 3 \pi\ov 6} (-1 + n^2) \e(1+3\e) \right)(x\bar x)^\e\right.\cr
&\left. + {\l^2\ov 2} \left({\pi^2\ov 6} (1 +  \e) (-1 + n^2)\right)(x\bar x)^{2\e}+...\right].}
$$

Computation of the anomalous dimension goes in exactly the same way as for the perturbing field:
$$
\eqalign{
\D_{n,n}^g &=\D_{n,n}-{\e\l\ov 2}\p_\l G_n(1,\l)=\cr
&=\D_{n,n}+{\sqrt 3\pi g\ov 12} \e^2 (1 + 3 \e) (-1 + n^2)-{\pi^2 g^2\ov 12} \e^2 (-1 + n^2),}
$$
where we again kept the appropriate terms of order $\e\sim g$. Then, at the fixed point \fx, this becomes
$$
\D_{n,n}^{g^*}=
{(-1 + n^2)( \e^2 + 3 \e^3 + 7 \e^4+...)\ov 8}
$$
which coincides with the dimension of the first component of the superfield $\Phi_{n,n}^{(p-2)}$
of the model $SM_{p-2}$.

\subsec{Matrix of anomalous dimensions}

Let us describe briefly the renormalization scheme of \pog\ . It
is a variation of that originally proposed by Zamolodchikov \zam\ .
The renormalized fields are expressed through the bare ones by:
$$
\phi^g_\a=B_{\a\b}(\l)\phi_\b
$$
(here $\phi$ could be the first or last component).
The two-point functions of the renormalized fields 
\eqn\norm{
G_{\a\b}^g(x)=<\phi_\a^g(x)\phi_\b^g(0)>,\quad G_{\a\b}^g(1)=\delta_{\a\b}, }
satisfy the equation
$$
(x\p_x-\b(g)\p_g)G_{\a\b}^g+\sum_{\rho=1}^2(\G_{\a\rho}G_{\rho\b}^g+\G_{\b\rho}G_{\a\rho}^g)=0
$$
where the matrix of anomalous dimensions $\Gamma$ is given by 
\eqn\ano{ \G=B\hat\D
B^{-1}-\e\l B\p_\l B^{-1} }
where $\hat\D=diag(\D_1,\D_2)$ is a
diagonal matrix of the bare dimensions. 
The matrix $B$ itself is
computed from the matrix of the bare two-point functions we computed
using the normalization condition \norm\ and requiring the matrix
$\G$ to be symmetric. Exact formulas can be found in \pog, here we
present our results for the supersymmetric case.

We computed above some of the entries of the $3\times 3$ matrix
of two-point functions in the second order. This matrix is obviously
symmetric. It turns out also that the remaining functions
$<\tphi_{n,n-2}(1)\tphi_{n,n-2}(0)>$ and
$<\phi_{n,n}(1)\tphi_{n,n-2}(0)>$ can be obtained from the computed
ones $<\tphi_{n,n+2}(1)\tphi_{n,n+2}(0)>$ and
$<\phi_{n,n}(1)\tphi_{n,n+2}(0)>$ by just taking $n\rightarrow -n$.
Let us denote for convenience the basis of fields:
$$
\eqalign{ \phi_1=\tphi_{n,n+2},\quad
\phi_2=(2\D_{n,n}(2\D_{n,n}+1))^{-1}\p\bar\p \phi_{n,n},\quad
\phi_3=\tphi_{n,n-2}, }
$$
where we normalized the field $\phi_2$ so that its bare two-point function is $1$. It is straightforward to
modify the functions involving $\phi_2$ taking into account the derivatives and the normalization.

We can write the matrix of the two-point functions up to the second
order in the perturbation expansion as:
$$
\eqalign{
G_{\a,\b}(x,\l)&=<\phi_\a(x)\phi_\b(0)>=
(x\bar x)^{-\D_\a-\D_\b}\left[\delta_{\a,\b}-\l C^{(1)}_{\a,\b}(x\bar x)^{\e}+{\l^2\ov 2}C^{(2)}_{\a,\b}(x\bar x)^{2\e}+...\right].}
$$
The two-point functions in the first order are proportional to the
structure constants \zam: \eqn\cfir{
 C^{(1)}_{\a,\b}=\hat C_{(1,3)(\a)(\b)}{\pi \g(\e+\D_\a-\D_\b)\g(\e-\D_\a+\D_\b)\ov
 \g(2\e)},}
which is ymmetric. Collecting all the dimensions and structure
constants, we get
$$
\eqalign{
C^{(1)}_{1,1}&=-{2 (3 + n) (-1 + 2 \e + \e n) \pi\ov \sqrt 3 \e (1 + n)},\quad
C^{(1)}_{1,2}={8 (-1 + \e) \sqrt{{2 + n\ov n}} \pi\ov \sqrt 3 \e (1 + n) (3 + n)},\quad
C^{(1)}_{1,3}=0,\cr
C^{(1)}_{2,2}&={8  \pi\ov \sqrt 3  (-1 + n^2) \e} - {4  (1 + n^2) \pi\ov
 \sqrt 3 (-1 + n^2) },\quad
C^{(1)}_{2,3}={8 (-1 + \e) \sqrt{{-2 + n\ov n}} \pi\ov \sqrt 3 \e (-3 + n) (-1 + n)},\cr
C^{(1)}_{3,3}&={-2 (-3 + n) (-1 + 2 \e - \e n) \pi\ov \sqrt 3 \e (-1 + n)}}
$$
for the first order, and
$$
\eqalign{
C^{(2)}_{1,1}&=-{2 (-20 - 143 n - 121 n^2 - 33 n^3 - 3 n^4) \pi^2\ov
 3 n (1 + n) (3 + n)^2 \e^2}-\cr
 &-{2 (5 + n) (8 + 151 n + 143 n^2 + 45 n^3 + 5 n^4) \pi^2\ov
 3 n (1 + n) (3 + n)^2 \e}\cr
C^{(2)}_{1,2}&=-{16 \sqrt{{2 + n\ov n}} (11 + 3 n) \pi^2\ov
  3 (1 + n) (3 + n) (5 + n) \e^2} +{ 16 \sqrt{{2 + n\ov n}} (57 + 18 n + n^2) \pi^2\ov
 3 (1 + n) (n+3)(n+5) \e }\cr
C^{(2)}_{1,3}&={80 (1 - 2 \e) \pi^2\ov 3 \e^2 n (-9 + n^2) \sqrt{-4 + n^2}}\cr
C^{(2)}_{2,2}&={32 \pi^2\ov 3 (-1 + n^2) \e^2} - {8 (19 + n^2) \pi^2\ov
 3 (-1 + n^2) \e}\cr
C^{(2)}_{2,3}&=-{16 \sqrt{{-2 + n\ov n}} (-11 + 3 n) \pi^2\ov
  3 (-1 + n) (-3 + n) (-5 + n)  \e^2} -  {16 \sqrt{{-2 + n\ov n}} (57 - 18 n + n^2) \pi^2\ov
 3 (-1 + n) (-3 + n) (-5 + n)  \e}\cr
C^{(2)}_{3,3}&= -{2 (-20 + 143 n - 121 n^2 + 33 n^3 - 3 n^4) \pi^2\ov
  3 n (-1 + n) (-3 + n)^2 \e^2} +\cr
  &+ {2 (-5 + n) (8 - 151 n + 143 n^2 - 45 n^3 + 5 n^4) \pi^2\ov
 3 n (-1 + n) (-3 + n)^2 \e} }
$$
for the second order.

Now we can apply the renormalization procedure of \pog\ and obtain
the matrix of anomalous dimensions \ano. Bare coupling constant $\l$ is
expressed through $g$ by \bare\ and the bare dimensions, up to order
$\e^2$. The results are:
$$
\eqalign{
\G_{1,1}&=\D_1-{(3 + n) (-1 + \e (2 + n)) \pi g\ov \sqrt 3 (1 + n)}+{4 g^2 \pi^2(2 + n)\ov 3 (1 + n)}\cr
\G_{1,2}&=\G_{2,1}=-{(-1 + \e) (-1 + n) \sqrt{{2 + n\ov 3n}} \pi g\ov (1 + n)}+{2 g^2 (-1 + n) \sqrt{{2 + n\ov n}} \pi^2\ov 3 (1 + n)}\cr
\G_{1,3}&=\G_{3,1}=0\cr
\G_{2,2}&=\D_2-{2\sqrt 3 \pi (-2 + \e  +  \e n^2 ) g\ov
 3 (-1 + n^2)}+{2 g^2 (3 + n^2) \pi^2\ov 3 (-1 + n^2)}\cr
\G_{2,3}&=\G_{3,2}=-{(-1 + \e) \sqrt{{-2 + n\ov 3n}} (1 + n) \pi g\ov (-1 + n)}\cr
 \G_{3,3}&=\D_3+{(1 + \e (-2 + n)) (-3 + n) \pi g\ov \sqrt 3 (-1 + n)}+{4 g^2 \pi^2(-2  + n )\ov 3 (-1 + n)},\ }
 $$
where
$$
\eqalign{
\D_1&=1 -{n+1\ov 2} \e + {1\ov 8} (-1 + n^2) \e^2,\quad
\D_2=1+{1\ov 8} (-1 + n^2) \e^2,\cr
\D_3&=1 +{n-1\ov 2} \e + {1\ov 8} (-1 + n^2) \e^2. }
$$
Evaluating this matrix at the fixed point \fx, we get
$$
\eqalign{
\G_{1,1}^{g^*}&=1 + {(20 - 4 n^2) \e\ov 8 (1 + n)} + {(39 - n - 7 n^2 + n^3) \e^2\ov
 8 (1 + n)}\cr
\G_{1,2}^{g^*}&=\G_{2,1}^{g^*}={(-1 + n) \sqrt{{2 + n\ov n}} \e(1+2\e)\ov n+1}\cr
\G_{1,3}^{g^*}&=\G_{3,1}^{g^*}=0\cr
\G_{2,2}^{g^*}&=1 + {4 \e\ov -1 + n^2} + {(65 - 2 n^2 + n^4) \e^2\ov 8 (-1 + n^2)}\cr
\G_{2,3}^{g^*}&=\G_{3,2}^{g^*}={\sqrt{{-2 + n\ov n}} (1 + n) \e(1+2\e)\ov n-1}\cr
\G_{3,3}^{g^*}&=1 + {(-5 + n^2) \e\ov 2 (-1 + n)} + {(-39 - n + 7 n^2 + n^3) \e^2\ov
 8 (-1 + n)} }
$$
whose eigenvalues are (up to order $\e^2$):
$$
\eqalign{
\D_1^{g^*}&=1 +  {1 + n\ov 2} \e + {7 +8 n + n^2\ov 8} \e^2\cr
\D_2^{g^*}&=1 + {-1 + n^2\ov 8} \e^2\cr
\D_3^{g^*}&=1 + {1-n\ov 2} \e +  {7 - 8 n + n^2\ov 8} \e^2. }
$$
This result coincides with dimensions $\D_{n+2,n}^{(p-2)}+1/2$,$\D_{n,n}^{(p-2)}+1$ and $\D_{n-2,n}^{(p-2)}+1/2$ of the model $SM_{p-2}$ up to this order.
The corresponding normalized eigenvectors should be identified with the fields of $SM_{p-2}$:
$$
\eqalign{ \tphi_{n+2,n}^{(p-2)}&={2 \ov n (1 + n)}\phi_1^{g^*} + {2
\sqrt{{2 + n\ov n}}\ov 1 + n}\phi_2^{g^*} + {\sqrt{-4 + n^2}\ov
n}\phi_3^{g^*}\cr \phi_2^{(p-2)}&=-{2 \sqrt{{2 + n\ov n}}\ov 1 +
n}\phi_1^{g^*} -{-5 +
  n^2\ov 1 + n^2}\phi_2^{g^*} +{2\sqrt{{n-2\ov n}}\ov n-1}\phi_3^{g^*}\cr
\tphi_{n-2,n}^{(p-2)}&={\sqrt{-4 + n^2}\ov n}\phi_1^{g^*}  - { 2
\sqrt{{-2 + n\ov n}}\ov n-1}\phi_2^{g^*} +{ 2\ov n(n-1)}\phi_3^{g^*}. }
$$
We used as before the notation $\tphi$ for the last component of the
corresponding superfield and:
$$
\phi_2^{(p-2)}={1\ov 2\D_{n,n}^{p-2}(2\D_{n,n}^{p-2}+1)}\p\bar\p
\phi_{n,n}^{(p-2)}
$$
is the normalized derivative of the corresponding first component. We notice that these eigenvectors are finite
as $\e\rightarrow 0$ with exactly the same combinations just as in (nonsupersymmetric) minimal models.

As we mentioned in the begining of this section the corresponding first components of $\Phi_{n,n\pm 2}$   and the last component of $\Phi_{n,n}$ will be also mixed along the RG flow in an analogous way.
This is thaks to the supersymmetry conserved by a perturbation with the last component of a superfield.
So we do not present a separate calculation for them.

\newsec{ Mixing of the fields in the Ramond sector}

\subsec{Conformal blocks in the Ramond sector}

The computation of the conformal blocks in the Ramond sector is more involved.
A way of computing them was recently proposed in \bm\ where
conformal blocks in the first few levels were shown to coincide with the
instanton partition function of certain $N=2$ YM theories in four
dimensions by a generalized AGT correspondence up to prefactors.

Following \bm\ one can compute NS-R conformal blocks only for a
special choice of the points. After that we can get the function
necessary for the integration in the second order by using its
conformal transformation properties.

The difficulties arise because of the branch cut in the OPE of
Ramond fields with the supercurrent:
\eqn\gr{ G(z)R^\ve(0)={\b
R^{-\ve}(0)\ov z^{{3\ov 2}}}+{G_{-1}R^\ve(0)\ov z^\hf}} where
$\b=\sqrt{\D-{\hat c\ov 16}}$, $\ve=\pm 1$.
Therefore one cannot obtain the usual commutation relations.
Here the Ramond field $R^\ve$ is doubly degenerate
because of the zero mode of $G$ in this sector.

The difficulty can be removed in the following way. Consider the OPE
between NS and Ramond fields: \eqn\rop{ \eqalign{
\phi_1(x)R_2^\ve(0)&=x^{\D-\D_1-\D_2}\sum_{N=0}^\infty x^N C_N^\ve
R_\D^\ve(0),\cr
\tilde\phi_1(x)R_2^\ve(0)&=x^{\D-\D_1-\D_2-\hf}\sum_{N=0}^\infty x^N
\tilde C_N^\ve R_\D^{-\ve}(0) }.}
Here $N$ runs over nonnegative
integers as $G$'s have integer valued modes in the Ramond sector.
Applying $G_0$ on both sides of \rop\ and taking into account \gr,
we obtain:
\eqn\gor{ \eqalign{ G_0C_N^\e &=\tilde
C_N^\ve+\b_2C_N^{-\ve},\cr G_0\tilde C_N^\ve
&=(\D-\D_2+N)C_N^\ve-\b_2\tilde C_N^{-\ve}.}}
From the consistency
conditions, $\tilde C_0^\ve$ is given by
$$
\tilde C_0^\ve=\b C_0^\ve-\b_2C_0^{-\ve}.
$$
Acting with $G_k$ with $k>0$ gives chain relations:
\eqn\rcr{ \eqalign{ G_k C_N^\e &=\tilde C_{N-k}^\e\cr G_k\tilde
C_N^\e &=(\D+2k\D_1-\D_2+N-k) C_{N-k}^\e }} and $L_k$ acts as usual
with the appropriate dimensions (see \bm\ for the details).

One has to solve these chain relations order by order or to use the recursion formulae. Then the conformal block for the function $<N(x)R(0)N(1)R(\infty)>$ ($N$ here stays for the first or the last component of a NS field)
 is obtained in the same way as in the NS case:
$$
F(x,\D,\D_i)=\sum_{N=0}^\infty x^N C_N(\D,\D_3,\D_4)S_N^{-1}C_N(\D,\D_1,\D_2)
$$
where $C_N$ could be actually $C_N$ or $\tilde C_N$ depending on the function in consideration.
Finally the correlation function is constructed as:
$$
<N(x)R(0)N(1)R(\infty)>=\sum_n C_n|F_n(x)|^2,
$$
where
$C_n$'s are the structure constants and the range of $n$ is dictated by the fusion rules.
The function that enters into the integral is then obtained by the conformal transformation.

As already mentioned, the conformal block in general is very
complicated. Fortunately, it is sufficient to compute the finite
term as $\e\rightarrow 0$. We did the computation for the functions
below up to high order and then check the behavior near the singular
points. It turns out also that the two-point function do not depend
on which of the fields $R^\ve$ are involved. So we drop the subscript $\ve$
from our notations in what follows.

\subsec{Function $<R_{n,n+1}(1)R_{n,n+1}(0)>$ }

Our calculation for the corresponding second order gives:
$$
\eqalign{ &<\tphi(x)R_{n,n+1}(0)R_{n,n+1}(\infty)\tphi(1)>={n^2-1\ov
12n^2}\left|{1\ov x (1 - x)^2} (1 + {n\ov n + 1} x - {1\ov n + 1}
x^2)\right|^2+\cr &+{(2 + n)^2\ov 48 n^2}\left|{1\ov x (1 - x)^2} (1 +  {2n\ov
n + 2} x + {n - 2\ov n + 2} x^2)\right|^2+{n+3\ov 12(n+1)}\left|{1\ov
(1-x)^2}(1+x)\right|^2. }
$$
To obtain the function that enters the integral, we use the conformal
transformation properties. One can easily get: \eqn\trsf{ \eqalign{
&<\tphi(x)R_{n,n+1}(0)R_{n,n+1}(1)\tphi(\infty)>=(x\bar
x)^{-2\D_{1,3}-1}<\tphi({x-1\ov
x})R_{n,n+1}(0)R_{n,n+1}(\infty)\tphi(1)>\cr &={n^2-1\ov
12n^2}\left|{(2x-1)(n x+1)\ov (n+1)x(x-1)}\right|^2+{(2 + n)^2\ov 48
n^2}\left|{(2x-1)(n(2x-1)+2)\ov (n+2)x(x-1)}\right|^2+{n+3\ov 12(n+1)}\left|{2x-1\ov
x}\right|^2. } }

We first integrate over the safe region, where $I(x)\sim
\pi/\e$. The result is:
$$
\eqalign{
&\int_{\O_{l,l_0}}I(x)<\tphi(x)R_{n,n+1}(0)R_{n,n+1}(1)\tphi(\infty)>=\cr
&={\pi^2\ov \e l^2}-{\pi^2 (20  + 13 n )\log l\ov 24 \e n}-{\pi^2 (4  + 5 n )\log{2l_0}\ov 24 \e n}-{\pi^2\ov 2\e}. }
$$
From this we have to subtract the lens-like region integral:
$$
{\pi^2(5n+4)\ov 24n\e}(\log l-\log{2l_0}).
$$

Next we proceed with the calculation of the integrals near the
singular points. Near $0$ (and near $1$ which  gives the exactly
same result) we use the OPE: \eqn\oper{ \eqalign{
\tphi(0)R_{n,n+1}(0)&=(x\bar x)^{-\D_{1,3}-1/2}\tilde
C_{(1,3)(n,n+1)}^{(n,n+1)}R_{n,n+1}(0)+\cr &+(x\bar
x)^{\D_{n,n-1}-\D_{n,n+1}-\D_{1,3}-1/2}\tilde
C_{(1,3)(n,n+1)}^{(n,n-1)}R_{n,n-1}(0)+\cr &+(x\bar
x)^{\D_{n,n+3}-\D_{n,n+1}-\D_{1,3}-1/2}\tilde
C_{(1,3)(n,n+1)}^{(n,n+3)}R_{n,n+3}(0). } }
We can approximate here $I(x)\sim \pi/\e-\pi\log|x|^2$, the necessary structure constants read:
\eqn\struc{\eqalign{
(\tilde C_{(1,3)(n,n+1)}^{(n,n+1)})^2&={-(2 + n)^2 (-1 + \e (-2 + 4 n))\ov 48 n^2}\cr
(\tilde C_{(1,3)(n,n+1)}^{(n,n-1)})^2&={(1 + 2 \e) (-1 + n^2)\ov 12 n^2}. }}
We remind again that all the structure constants involving the field $\tphi$ should be divided by $2\D_{1,3}$ and we keep
in what follows the same notation $\tilde C$.

Then, the result of the integration is:
$$
\eqalign{
&\int_{D_{l,0}\backslash D_{l_0,0}}I(x)<\tphi(x)R_{n,n+1}(0)R_{n,n+1}(1)\tphi(\infty)>=\cr
&={\pi^2 (28 + 40 n + 12 n^2 + n^3)\ov 24 \e^2 n (2 + n)^2}-{
   \pi^2  (4 + 24 n + 36 n^2 + 15 n^3 + 2 n^4)\ov 12 \e n (2 + n)^2}+{\pi^2(4 + 5 n)\log l\ov 24 \e n}. }
$$

Around $\infty$, we make the transformation $x\rightarrow
1/x$ and then $x\rightarrow 0$ as usual. The structure constant is:
$$
\hat C_{(1,3)(1,3)}^{(1,3)}\tilde C_{(1,3)(n,n+1)}^{(n,n+1)}={(2 + n) (1 -2 \e - 2 \e n) \ov 6  n}
$$
and we found:
$$
\eqalign{
&\int_{D_{l,\infty}\ D_{l_0,\infty}}I(x)<\tphi(x)R_{n,n+1}(0)R_{n,n+1}(1)\tphi(\infty)>=\cr
&=-{ \pi^2\ov l^2 \e}+{(2 + n) \pi^2\ov 6 n \e^2}-{(2 + n) \pi^2 (1 + n - \log l)\ov 3 n \e.} }
$$
Collecting all the terms, we obtain:
$$
{\pi^2\left[44 + 64 n + 24 n^2 + 3 n^3 -
   8 \e (1 + n) (5 + 14 n + 7 n^2 + n^3)\right]\ov 12 \e^2 n (2 + n)^2}.
$$

\subsec{Function $<R_{n,n-1}(1)R_{n,n+1}(0)>$ }

The calculation of the four-point function with the perturbing
fields can be done in the same way:
$$
<\tphi(x)R_{n,n+1}(0)\tphi(1)R_{n,n-1}(\infty)>={\sqrt{n^2-1}\ov 12
n}\left|{1\ov x(1-x)}(1+x)\right|^2.
$$
Performing the same transformation as in \trsf, the integrand becomes:
$$
<\tphi(x)R_{n,n+1}(0)R_{n,n-1}(1)\tphi(\infty)>={\sqrt{n^2-1}\ov 12 n}\left|{2x-1\ov x(1-x)}\right|^2.
$$
This function is almost the same as in the nonsupersymmetric case but we should
calculate again because the various structure constants
and dimensions of the fields are different. The
integration over the safe region and lens-like regions are exactly
the same and the results are, respectively,
$$
{8 \sqrt{n^2-1}\pi^2 (5 \log l + \log{2l_0})\ov 3n\e (n^2-16)},\quad
-{8 \sqrt{n^2-1}\pi^2 ( \log l - \log{2l_0})\ov 3n\e (n^2-16)}.
$$

For the calculation around $0$ we use the same OPE that appeared in the previous subsection 6.6
(without the last line because of the fusion rules).
In addition to the structure constants presented above, we need also:
$$
\tilde C_{(1,3)(n,n-1)}^{(n,n-1)}={n-2\ov 4 \sqrt 3 n} [1 + (2 n+1) \e].
$$
The result is:
$$
\eqalign{
&\int_{D_{l,0}\backslash D_{l_0,0}}I(x)<\tphi(x)R_{n,n+1}(0)R_{n,n-1}(1)\tphi(\infty)>=\cr
&=-{2 \sqrt{n^2-1} \pi^2 [-28 + n^2 + 2 \e (4 + 5 n^2) +
    4 \e (n^2-4)\log l ]\ov 3n \e^2 (64 - 20 n^2 + n^4)}. }
$$
The integral around $1$ gives in the same result as around $0$. 
The integral around $\infty$ with the structure constants \struc\ gives:
$$
\eqalign{
\int_{D_{l,\infty}\backslash D_{l_0,\infty}}I(x)<\tphi(x)R_{n,n+1}(0)R_{n,n-1}(1)\tphi(\infty)>={16\sqrt{n^2-1}
\pi^2 (-1 + 2 \e - 2 \e \log l)\ov 3n \e^2 (n^2-16)}. }
$$
We collect now all the terms and obtain the final result in the second order:
$$
{4\pi^2 \sqrt{n^2 - 1} (44 - 5 n^2-2\e (20+n^2)) \ov 3 \e^2 n(n^2-16)(n^2-4)}.
$$

\subsec{Matrix of anomalous dimensions}

The functions we computed above are enough for our computation since the other two functions
$<R_{n,n-1}(1)R_{n,n-1}(0)>$ and $<R_{n,n+1}(1)R_{n,n-1}(0)>$ can be
obtained from $<R_{n,n+1}(1)R_{n,n+1}(0)>$ and
$<R_{n,n-1}(1)R_{n,n+1}(0)>$ by just changing $n\rightarrow -n$  as in the
case of NS fields.
Let us introduce again a basis: $R_1=R_{n,n+1}$, $R_2=R_{n,n-1}$. From
the general formula \cfir\ and the bare dimensions of the fields
$$
\eqalign{
\D_1&={3\ov 16}  -\left({n\ov4}+{1\ov8}\right) \e + {1\ov 8} (n^2-1) \e^2,\cr
\D_2&={3\ov 16}  +\left({n\ov4}-{1\ov8}\right) \e + {1\ov 8} (n^2-1) \e^2,  }
$$
we get for the $2\times 2$ matrix of two-point functions in the first order:
$$
\eqalign{
C^{(1)}_{1,1}&={(2 + n) \pi\ov 2 \sqrt 3 n \e} - {(2 + n) (-1 + 2 n)\pi\ov
 2 \sqrt 3 n},\cr
C^{(1)}_{1,2}&=C^{(1)}_{2,1}=-{4\sqrt{n^2-1}\pi(1+\e)\ov\sqrt 3 n (n^2-4) \e },\cr
C^{(1)}_{2,2}&={(n-2) \pi\ov 2 \sqrt 3 n \e} + {(n-2) (2 n+1)\pi\ov
 2 \sqrt 3 n}. }
$$
and, in the second order
$$
\eqalign{
C^{(2)}_{1,1}&={(44 + 64 n + 24 n^2 + 3 n^3) \pi^2\ov 12 \e^2 n (n+2)^2} -
{ 2 (n+1) (5 + 14 n + 7 n^2 + n^3) \pi^2\ov 3 \e n (2 + n)^2},\cr
C^{(2)}_{1,2}&=C^{(2)}_{2,1}={4 \sqrt{n^2 - 1} (44 - 5 n^2) \pi^2\ov
 3 \e^2  n (n^2-1)(n^2-4)} - {
 8 \sqrt{n^2 - 1}  (n^2+20) \pi^2\ov 3 \e n (n^2-16)(n^2-4)},\cr
C^{(2)}_{2,2}&={(-44 + 64 n - 24 n^2 + 3 n^3)\pi^2\ov 12 \e^2 (n-2)^2 n}+
{ 2 (n-1) (-5 + 14 n - 7 n^2 + n^3) \pi^2\ov 3 \e (n-2)^2 n}. }
$$

Now, following the same procedure as in the NS case, we get for the matrix of anomalous dimensions up to order
$\e^2\sim g^2$:
$$
\eqalign{
\G_{1,1}&=\D_1-{g (n+2) (-1 - \e + 2 \e n) \pi\ov 4 \sqrt 3 n}+{g^2 \pi^2\ov 4},\cr
\G_{1,2}&=\G_{2,1}={(1 + \e) g \sqrt{n^2-1} \pi\ov 2 \sqrt 3 n},\cr
\G_{2,2}&=\D_2+{g (n-2) (1 + \e + 2 \e n) \pi\ov 4 \sqrt 3 n}+{g^2 \pi^2\ov 4}, }
$$
which becomes at the fixed point \fx:
$$
\eqalign{
\G_{1,1}^{g^*}&={3\ov 16} + {(4 + n - 2 n^2) \e\ov 8 n} + {(8 + n - 4 n^2 + n^3) \e^2\ov 8 n},\cr
\G_{1,2}^{g^*}&=\G_{2,1}^{g^*}={\sqrt{n^2-1} \e(1+2\e)\ov 2 n},\cr
\G_{2,2}^{g^*}&={3\ov 16} + {(-4 + n + 2 n^2) \e\ov 8 n} + {(-8 + n + 4 n^2 + n^3) \e^2\ov 8 n}. }
$$
The eigenvalues of this matrix are
$$
\eqalign{
\D_1^{g^*}&={3\ov 16} + \left({1\ov 8} + {n\ov 4}\right) \e + {1\ov 8} (1 + 4 n + n^2) \e^2,\cr
\D_2^{g^*}&={3\ov 16} + \left({1\ov 8} - {n\ov 4}\right) \e + {1\ov 8} (1 - 4 n + n^2) \e^2. }
$$

As expected, they coincide with the dimensions of the Ramond fields $\D_{n+1,n}^{(p-2)}$ and $\D_{n-1,n}^{(p-2)}$ of the  $SM_{p-2}$. The corresponding fields are expressed as a  (normalized) linear combination:
$$
\eqalign{
R_{n+1,n}^{(p-2)}&={1\ov n}R_1^{g^*}+{\sqrt{n^2-1}\ov n}R_2^{g^*},\cr
R_{n-1,n}^{(p-2)}&=-{\sqrt{n^2-1}\ov n}R_1^{g^*}+{1\ov n}R_2^{g^*}. }
$$

\newsec{Concluding remarks}

To conclude, we considered here the the RG flow of the minimal superconformal model $SM_p$ with $p\gg 1$ up to the second order in the perturbation theory. It is confirmed that there is a nontrivial fixed point that coincides with the model $SM_{p-2}$, which was established before up to the first order calculations. 
We computed the anomalous dimensions of certain fields in both NS and Ramond sectors along the RG flow. At the fixed point they coincide with the dimensions of the corresponding fields from $SM_{p-2}$.

We would like to make two comments at the end. Firstly, we have found that
the linear combinations (i.e. the eigenvectors of the matrix of anomalous dimensions) expressing the fields in the $SM_{p-2}$ 
do not depend on $\e$ in both the NS and Ramond sectors. This happens also in the non-supersymmetric case. 
So one can speculate that they are actually exact.
Secondly, the coefficients in this linear combination are exactly the same. 
This may suggest that the same linear combination becomes the eigenvectors for all the $SU(2)$
coset theories perurbed by the least relevant field.

\

\noindent
{\bf Acknowledgements}

This work was supported in part by the WCU Grant No. R32-2008-000-101300 and the Research fund no. 1-2008-2935-001-2
 by Ewha Womans University (CA).
\listrefs

\bye